\RequirePackage{fix-cm}
\documentclass[natbib,twocolumn,final]{svjour3}          % twocolumn
\usepackage{soul}
\usepackage{amssymb}
\usepackage{hyperref}
\usepackage{mathtools, cuted}
\usepackage{commath}
\usepackage{lineno}
\usepackage{siunitx}
\usepackage{graphicx}
\usepackage{amssymb,amsmath}
\usepackage{natbib}
\usepackage{aps-bibstyle}  % use this style if you upload to .tex file only a part of Bibtex created bbl.
\usepackage[first=0,last=9]{lcg}
\usepackage{url}
 \journalname{to be inserted}
\newcommand{\aap}{{Astron. Astrophys.}}

\begin{document}
%%%%%%%%%%%%%%%%%%%%%%%%%%%%%%%%%%%%%%%%%%%%%%%%%%%%%%%%%%%%%%%%%%%%%%%%
\title{The perturbed restricted three-body problem with angular velocity: Analysis of basins of convergence linked to the libration points}
%\subtitle{The analysis of periodic orbit in the restricted four-body problem using mobile coordinates}
%\titlerunning{The perturbed R with angular velocity: Analysis of basins of convergence linked to the libration points}% if too long for running head
%\shortauthors{Suraj et al.}
\author{Md Sanam Suraj  \and
        Rajiv Aggarwal       \and
              Amit Mittal\and
              Md Chand Asique
}
\institute{Md Sanam Suraj \at
    Department of Mathematics,
    Sri Aurobindo College, University of Delhi,  New Delhi-110017, Delhi, India\\
    \email{\url{mdsanamsuraj@gmail.com}}\\
    \email{\url{mdsanamsuraj@aurobindo.du.ac.in}}           %  \\
           \and
  Rajiv Aggarwal \at
  Department of Mathematics,
  Deshbandhu College, University of Delhi, New Delhi-110019, Delhi, India\\
              \email{\url{rajiv_agg1973@yahoo.com}}
%%%%%%%%%%%%%%%%%%%%%%%%%%%%%%%%%%%%%%%%%%%%%%%%%%%%%
\and
Amit Mittal\at
Department of Mathematics,
ARSD College, University of Delhi, New Delhi-110021, Delhi,  India\\
 \email{\url{to.amitmittal@gmail.com}}
 %%%%%%%%%%%%%%%%%%%%%%%%%%%%%%%%%%%%%%%%%%%%%%%%%%%%%
\and
Md Chand Asique\at
Deshbandhu College, University of Delhi, New Delhi-110019, Delhi,  India\\
 \email{\url{mdchandasique@gmail.com}}
 }
\date{Received:     date / Accepted:             date}
%\date{today}
\maketitle
\begin{abstract}
The analysis of the affect of angular velocity on the geometry of the basins of convergence (BoC) linked to the equilibrium  points in the  restricted three-body problem is illustrated when the primaries are source of radiation. The bivariate scheme of the Newton-Raphson (N-R) iterative method has been used to discuss the  topology of the basins of convergence. The parametric evolution of the fractality of the convergence plane is also presented where the degree of fractality is illustrated by evaluating the basin entropy of the convergence plane..
\end{abstract}
%%%%%%%%%%%%%%%%%%%%%%%%%%%%%%%%%%%%%%%%%%%%%%%%%%%%%%%%%%%%%%%%%%%%%%%%
%%%%%%%%%%%%%%%%%%%%%%%%%%%%%%%%%%%%%%%%%%%%%%%%%%%%%%%%%%%%%%%%%%%%%%%%%%%
%\phantomsection
%\addcontentsline{toc}{section}{Introduction}
\section{Introduction}
\label{intro:1}
%%%
One of the most celebrated problem in the field of Celestial Mechanics is the restricted three-body problem (R3BP). Many researchers and scientists are attracted towards it due to its applications in various other fields (e.g. \cite{AA19a}, \cite{AGL19b}, \cite{A12}, \cite{AAG09}, \cite{AAG17}, \cite{CM87}, \cite{S07}, \cite{PAT19}, \cite{SGA19}). In addition, several modifications have been proposed by various researchers to be more realistic in the classical R3BP which make the applications of this problem in wider sense.  In this proposed problem we have considered two modifications i.e., the radiation effects of both the primaries and the variation in the angular velocity (Chermnykh problem,  see \cite{C87}).  A generalization of the Euler's problem  of two fixed gravitational centers and the restricted problem of three bodies where the third body, whose mass is negligible in comparison of the other bodies, orbits in the configuration plane of dumbbell which rotates around their center of mass with a constant angular velocity $\omega$,  is always referred as Chermnykh problem. Many authors have studied this problem due to its important applications in the field of Chemistry (\cite{PFG96}), Celestial mechanics and Dynamical Astronomy.

One of the paramount issue in the dynamical system is to know the geometry of the "basins of convergence" linked with the equilibrium points of the dynamical system.  The domain of the BoC unveils the fact that how the different initial conditions on the configuration  plane are enticed by the particular equilibrium point when an iterative method is applied to solve the system of equations. Undoubtedly, to solve the system of simultaneous equations with two or more variables, the N-R iterative scheme is contemplated as classical one. Previously, many authors have studied the BoC by applying the N-R iterative method to reveal the numerous intrinsic properties of different dynamical system (e.g., the R3BP  and the Hill's problem with oblateness and radiation effects(\cite{Z16}, \cite{Z17}, \cite{D10}), the restricted four-body problem (\cite{BP11},\cite{SAP17}, \cite{SAA17}, \cite{SUR20}), the restricted problem of five bodies (\cite{ZS18}, \cite{SAR19}, \cite{sur19}, \cite{sur19b}, \cite{Sur19d}).

Presently, we wish to analyze the effect of angular velocity on the topology of the BoC when both of the primaries are source of radiation. Moreover, the fractality of the BoC is also discussed as the function of angular velocity.

The present paper has following structure: along with the literature review regarding the R3BP presented in Sec. \ref{intro:1},  the description  of the mathematical model is presented in Sec. \ref{Sec:2}. The parametric evolution of the locations of the equilibrium points is depicted in Sec. \ref{Sec:3} whereas the influences of the angular velocity on the geometry of the BoC  by using the bivariate sort of the N-R iterative method are illustrated in detail in Sec. \ref{Sec:4}.  The degree of fractality of the BoC is depicted in Sec. \ref{Sec:5}.  The paper ends with Sec. \ref{Sec:6} where the analysis of the study and the obtained results are  discussed.
%%%%%%%%%%%%%%%%%%%%%%%%%%%%%%%%%%%%%%%%%
%%%%%%%%%%%%%%%%%%%%%%%%%%%%%%%%%%%%%%%%%
\section{Mathematical descriptions and the equations of motion}
\label{Sec:2}
\begin{figure}
\centering
\resizebox{\hsize}{!}{\includegraphics{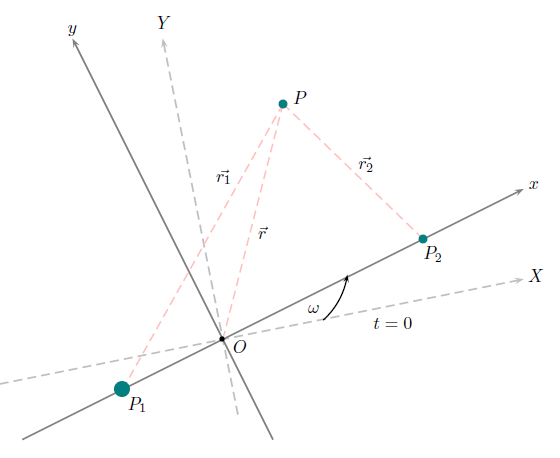}}
\caption{The restricted three-body problem.   (colour figure online).}
\label{Fig:00}
\end{figure}

In the present study, we have considered the dynamical model same as in Ref.\cite{PK15} which can be reviewed as follows: the rotating, barycentric, and a dimensionless co-ordinate system with origin "O" is considered as the centre of mass of the system. The two primaries namely $m_1$ and $m_2$ rotate in circular orbit around "O" with angular velocity $\omega \geq 0$ and in addition, the primaries always lie on the $x-$axis with co-ordinate $(x_1, 0)=(-\mu, 0)$ and $(x_2, 0)=(1-\mu, 0)$ (see Fig. \ref{Fig:00}). In dimensionless unit $m_1=1-\mu$ and $m_2=\mu$ where the mass parameter $\mu=\frac{m_2}{m_1+m_2} \leq \frac{1}{2}$. The restricted problem of three bodies reduces to the Copenhagen problem when $\mu=\frac{1}{2}$.  We analyse the motion of the third body $m_3$ whose mass is negligible in comparison of the primaries. In addition, it is also considered that both the primaries are source of radiation. Consequently, the motion of the infinitesimal mass is governed by two type of forces i.e., the gravitational forces of the primaries and the repulsive force of the light pressure. It is necessary to note that the radiation factors can achieve the negative value as well which means that these forces will give strength to the gravitational force.

In the dimensionless rectangular rotating  co-ordinate system, the equations of motion of the third body, which also referred as test particle in the restricted three-body problem with angular velocity, are (see \cite{C87},  \cite{PR04} and \cite{PK15}):
%%%%%%%%%%%%%%%%%%%%%%%%%%%%%%%%%%%%%%%%
\begin{subequations}
\begin{eqnarray}
\label{Eq:1a}
  \ddot{x} -2\dot{y}&=&(\omega^2-\mathfrak{Q}^*)x-\mathfrak{M}^*\mathfrak{R}^*,\\
  \label{Eq:1b}
  \ddot{y}+2\dot{x} &=&(\omega^2-\mathfrak{Q}^*)y,
\end{eqnarray}
\end{subequations}
where
\begin{subequations}
\begin{eqnarray}
\label{Eq:2a}
\mathfrak{M}^*&=&\mu(1-\mu),\\
\mathfrak{Q}^* &=& \frac{q_1(1-\mu)}{r_1^3}+\frac{q_2\mu}{r_2^3},\\
  \label{Eq:2b}
\mathfrak{R}^* &=&\frac{q_1}{r_1^3}-\frac{q_2}{r_2^3},
\end{eqnarray}
\end{subequations}
%%%%%%%%%%%%%%%%%%%%%%%%%%%%%%%%%%%%%%%%%%%%%
%%%%%%%%%%%%%%%%%%%%%%%%%%%%%%%%%%%%%%%%%%%%%
while  the time independent potential function $\Omega$ is given by:
\begin{subequations}
\begin{eqnarray}
\label{Eq:3a}
\Omega&=& \frac{\omega^2}{2}(x^2+y^2)+\sum_{i=1}^{2}\frac{q_i m_i}{r_i},\\
\label{Eq:3b}
 r_i^2 &=& \tilde{x_i}^2+\tilde{y_i}^2,\\
 \label{Eq:3c}
 \tilde{x_i}&=&(x-x_i),\\
  \label{Eq:3d}
 \tilde{y_i}&=&(y-y_i),
  \label{Eq:2e}
\end{eqnarray}
\end{subequations}
where $r_i$ represents the distances of the test particle from the primaries $m_i$, respectively. The radiation parameters $q_i$,(see \cite{C70}) due to the radiating primaries $m_i$ are defined as:
\begin{equation*}
  q_i=1-\frac{F_{p_i}}{F_{g_i}},
\end{equation*}
where, $F_{p_i}$ are the solar radiation pressure forces whereas $F_{g_i}$ are the gravitational forces due to primaries $m_i, i=1,2$.
The system admits the Jacobi integral i.e.,
\begin{equation}\label{Eq:4}
 C=2\Omega -(\dot{x}^2+\dot{y}^2).
\end{equation}
%%%%%%%%%%%%%%%%%%%%%%%%%%%%%%%%%%%%%%%%%%%%%
%%%%%%%%%%%%%%%%%%%%%%%%%%%%%%%%%%%%%%%%%%%%%
\begin{figure}
\centering
\resizebox{\hsize}{!}{\includegraphics{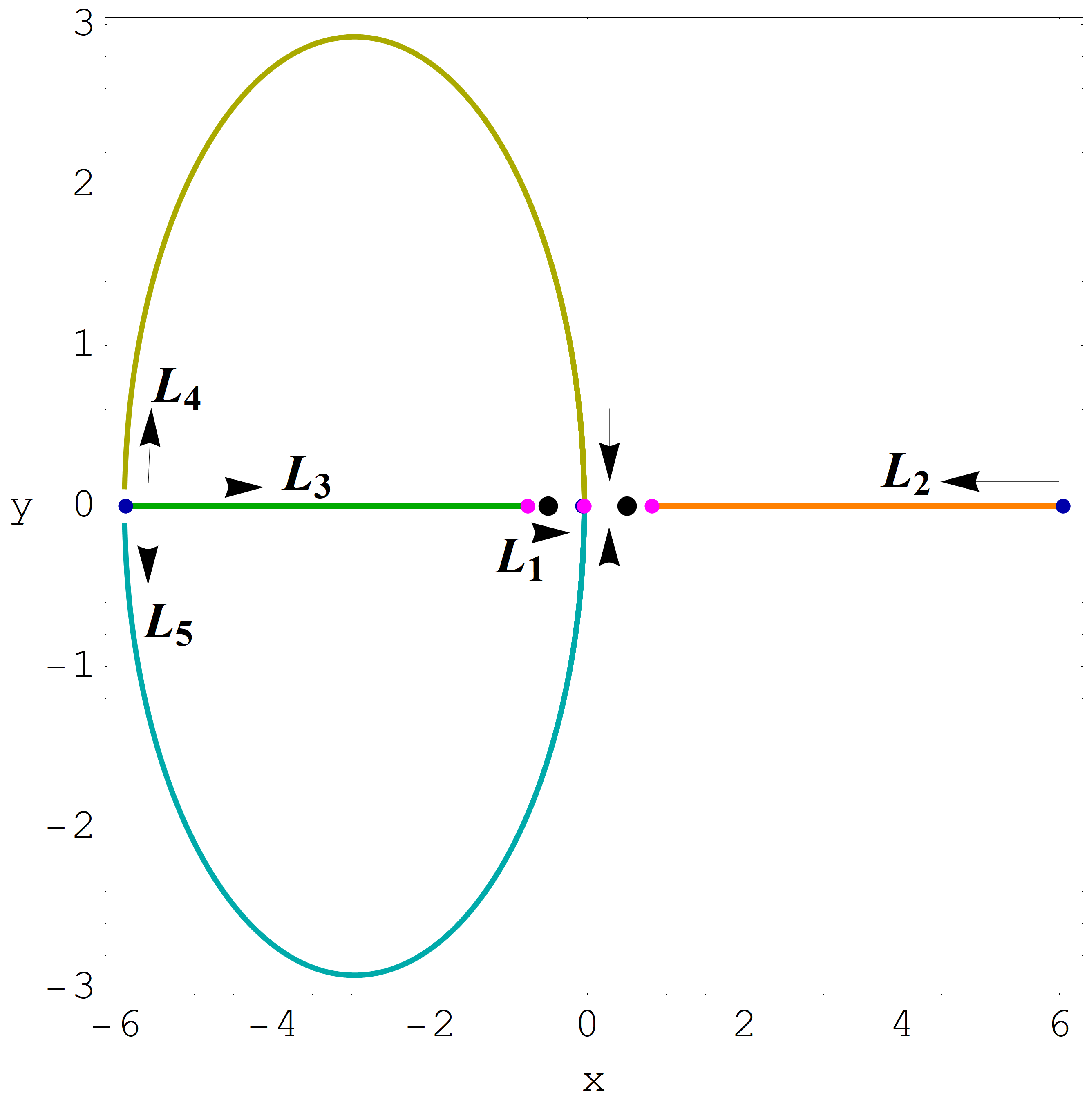}}
\caption{The movement of the libration points for $q_1=0.15$, $q_2=0.25$ and consequently $\omega \in (0.03097574, 1.25144256)$. The color codes are :  $L_1= $\emph{Purple}, $L_2= $\emph{Orange}, $L_3= $\emph{Green}, $L_4= $\emph{Olive}, $L_5= $\emph{Cyan}.          (colour figure online).}
\label{Fig:M1}
\end{figure}
%%%%%%%%%%%%%%%%%%%%%%%%%%%%%%%%%%%%%%%%%%%%%
%%%%%%%%%%%%%%%%%%%%%%%%%%%%%%%%%%%%%%%%%%%%%
\section{The libration points: a parametric evolution}
\label{Sec:3}
The parametric evolution of the positions of libration points are presented in this section by using the same procedure given by Ref.\cite{PK15}. The collinear libration points are those points which lie on $x-$axis and we can evaluate  by system of equations (\ref{Eq:1a}-\ref{Eq:1b}), by setting the velocity and acceleration components equal to zero and solving for $x$ by taking $y=0$, we get
\begin{equation}
\label{Eq: 5}
 f(x)=\omega^2x-\frac{(1-\mu)q_1(x+\mu)}{|x+\mu|^3}-\frac{\mu q_2(x+\mu-1)}{|x+\mu-1|^3}=0,\\
\end{equation}
by keeping the value of the  parameters $\omega, \mu$ and $q_i$, $i=1,2$, fixed. The presented problem reduces to the photo-gravitational version of the classical restricted problem when $\omega=1$.  It is shown that the angular velocity $\omega$ has no affect on the existence of totality of number of collinear libration points  (for detail see Ref.\cite{PK15}) and these libration points are named as $L_{i}$, $i=1,2,3$ where their positions are defined as follows:
\begin{align*}
  L_{3}< & -\mu<L_{1}<1-\mu<L_{2},
\end{align*}
where $-\mu$ and $1-\mu$ are the positions of the primaries $m_1$ and $m_2$ respectively.

As far as the non-collinear triangular libration points are concerned, their positions can be described as follows:
\begin{subequations}
\begin{eqnarray}
\label{Eq:6a}
  x &=& \frac{1}{2}\Big\{1+\Big(\frac{q_1}{\omega^2}\Big)^\frac{2}{3}-\Big(\frac{q_2}{\omega^2}\Big)^\frac{2}{3}\Big\}-\mu,\\
\label{Eq:6a}
  y&=&\pm\Big[\Big(\frac{q_1}{\omega^2}\Big)^\frac{2}{3}-\frac{1}{4}\Big\{1+\Big(\frac{q_1}{\omega^2}\Big)^\frac{2}{3}-\Big(\frac{q_2}{\omega^2}\Big)^\frac{2}{3}\Big\}^2\Big]^\frac{1}{2},\nonumber\\
\end{eqnarray}
\end{subequations}
for detail see Ref. \cite{PK15}. In addition,  it is unveiled that the planar non-collinear libration  points i.e., $y\neq 0$, exist only when the following conditions are satisfied simultaneously:
\begin{align}\label{Eq:7}
 q_1^\frac{1}{3} > 0, & \quad q_2^\frac{1}{3} > 0,  \text{and} \quad| q_1^\frac{1}{3}-q_2^\frac{1}{3}|<\omega^\frac{2}{3}<(q_1^\frac{1}{3}+q_2^\frac{1}{3}).
\end{align}
When the effect of the radiation pressure is neglected the positions of the non-collinear  equilibrium points are defined by the co-ordinates $(x , y)$ (see Ref. \cite{PK15}, \cite{PR04}) where
\begin{align}\label{Eq:8}
  x=\frac{1}{2}(1-2\mu), \quad  y=\pm \sqrt{\omega^{-\frac{4}{3}}-\frac{1}{4}},
\end{align}
and consequently, these points exist only when $\omega \in (0, 2\sqrt{2})$. It is necessary to mention that the non-collinear libration points exist only for the particular value of $\omega$ which depend on $q_1$ and $q_2$ (where $q_1, q_2\neq 1$), the triangular libration points exist only when $\omega\in(\omega_1, \omega_2)$ where $\omega_i=\omega_i(q_1, q_2), i=1,2$. However, the collinear libration points exist for $\omega \in (0, \infty)$ and at $\omega=2\sqrt{2}$(where $q_1, q_2=1$)  the non-collinear libration points coincide with $L_{1}$. In Fig. \ref{Fig:M1}, the movements of the position of libration points (as the value of parameter $\omega \in (\omega_1,\omega_2)$) are shown for constant values of the parameters $q_i$, and $\mu$ and different increasing values of $\omega$. We can observe that the libration point $L_{3}$ move towards the primary $P_1$ whereas the libration points $L_{1,2}$ move towards the primary $P_2$ as the value of $\omega$ increases. It is also observed that the non-collinear libration points originate in the vicinity of the libration point $L_{3}$ at $\omega\approx0.03097574$ and these points annihilate in  vicinity of the libration point $L_{1}$ at $\omega\approx1.25144256$. Moreover, these particular values of $\omega$ are associated to the value of $q_1=0.15$ and $q_2=0.25$.
%%%%%%%%%%%%%%%%%%%%%%%%%%%%%%%%%%%%%%%%%%%%%
\begin{figure*}
\centering
\resizebox{\hsize}{!}{\includegraphics{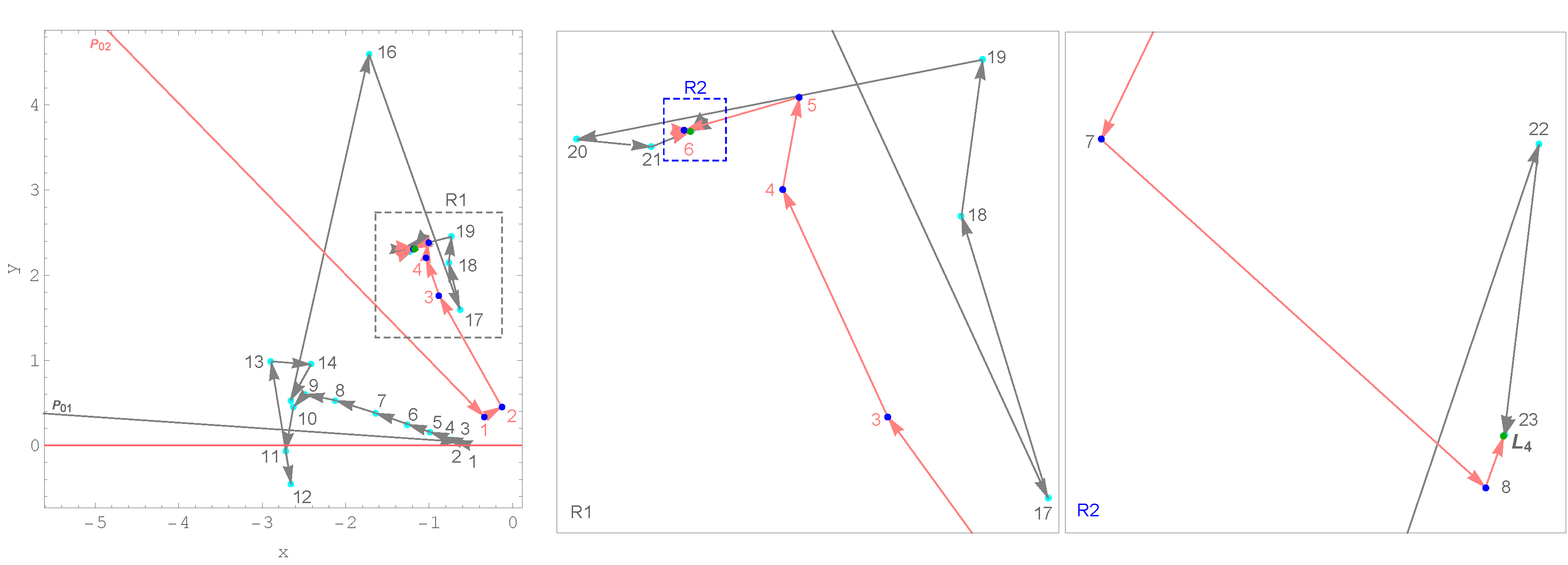}}
\caption{A characteristic example of the consecutive steps that are followed by the Newton-Raphson iterator and the corresponding crooked path-line that leads to an equilibrium point.   (colour figure online).}
\label{Fig:C1}
\end{figure*}
%%%%%%%%%%%%%%%%%%%%%%%%%%%%%%%%%%%%%%%%%%%%%
\section{The Newton-Raphson basins of convergence (N-RBoC)}
\label{Sec:4}
We perform a numerical analysis of the influence of angular velocity, mass parameter, radiation parameters on the geometry of the BoC linked with the libration points of the dynamical system by using the bivariate version of the N-R iterative scheme. This iterative method can be applicable to the of system of bivariate function $\mathbf{f(x)}=0$, using the iterative method:
\begin{equation}\label{Eq:}
 \mathbf{ x}_{n+1}=\mathbf{x}_n-\mathbf{J}^{-1}\mathbf{f(x_n)}.
\end{equation}
Here,   $\mathbf{f(x_n)}$ denotes the system of equations,  whereas $\mathbf{J}^{-1}$ is denoting the inverse Jacobian matrix.

The iterative scheme for the $x$ and $y$ co-ordinates can be decomposed as:
\begin{eqnarray*}
%\label{Eq:304a}
x_{n+1}&=&x_n-\frac{\Omega_{x_n}\Omega_{y_ny_n}-\Omega_{y_n}\Omega_{x_ny_n}}{\Omega_{x_nx_n}\Omega_{y_ny_n}-\Omega_{x_ny_n}\Omega_{y_nx_n}},\\
%\label{Eq:304b}
y_{n+1}&=&y_n+\frac{\Omega_{x_n}\Omega_{y_nx_n}-\Omega_{y_n}\Omega_{x_nx_n}}{\Omega_{x_nx_n}\Omega_{y_ny_n}-\Omega_{x_ny_n}\Omega_{y_nx_n}},
\end{eqnarray*}
where the values of the $x$ and $y$ coordinates are represented by $x_n$ and $y_n$ respectively at the $n$-th step.

The philosophy which works in the background of the N-R iterative scheme is same as described in \cite{Z16}.  The collection of all those initial conditions which converge to the particular attractor (i.e., the same root of the equations) compose the so-called N-RBoC. Further, we apply color coded diagrams (CCDs), where each pixel is linked with a non-identical color, as per the concluding state of the associated initial conditions, to classify the nodes in the orbital plane. The color codes for the domain of BoC linked to the respective libration points are same in each figure and the codes are same as in Fig. \ref{Fig:M1}. In Fig. \ref{Fig:C1}, it is depicted that the successive approximation points move in a crooked path and also for different initial conditions but for same attractor the number of required iterations to converge are different.
\begin{figure*}%
\centering
\includegraphics[scale=0.45]{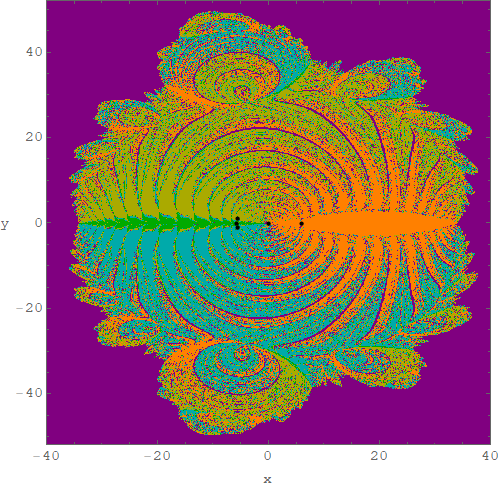}
\includegraphics[scale=0.45]{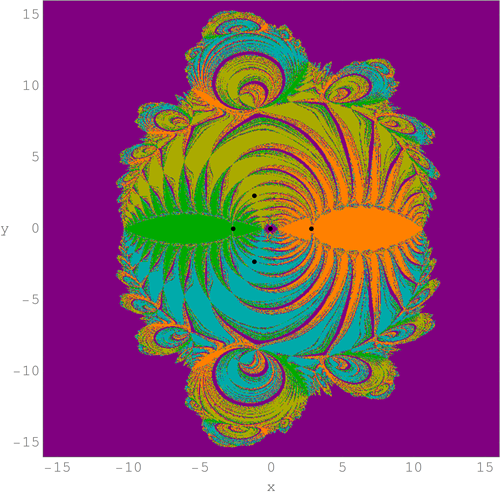}\\
\includegraphics[scale=0.18]{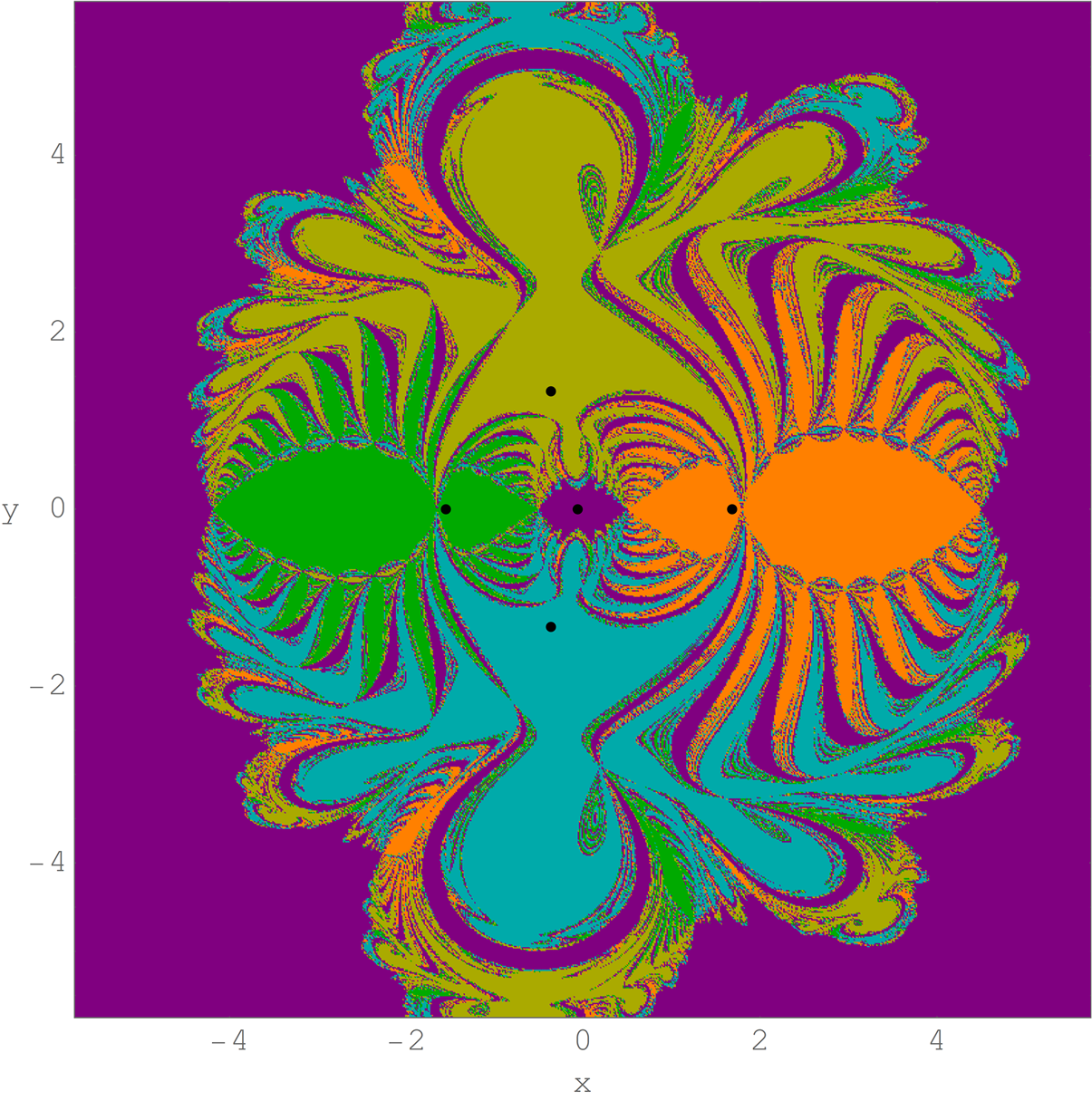}
\includegraphics[scale=0.435]{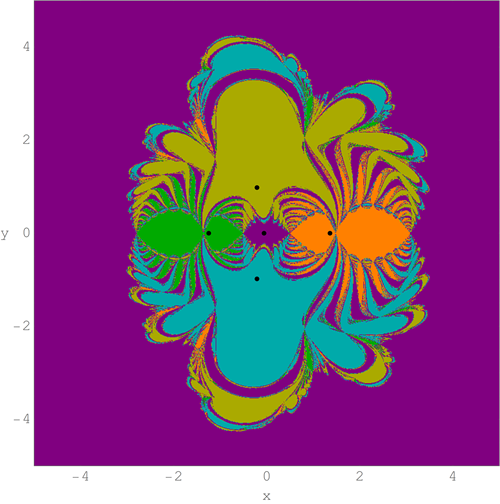}\\
\includegraphics[scale=0.18]{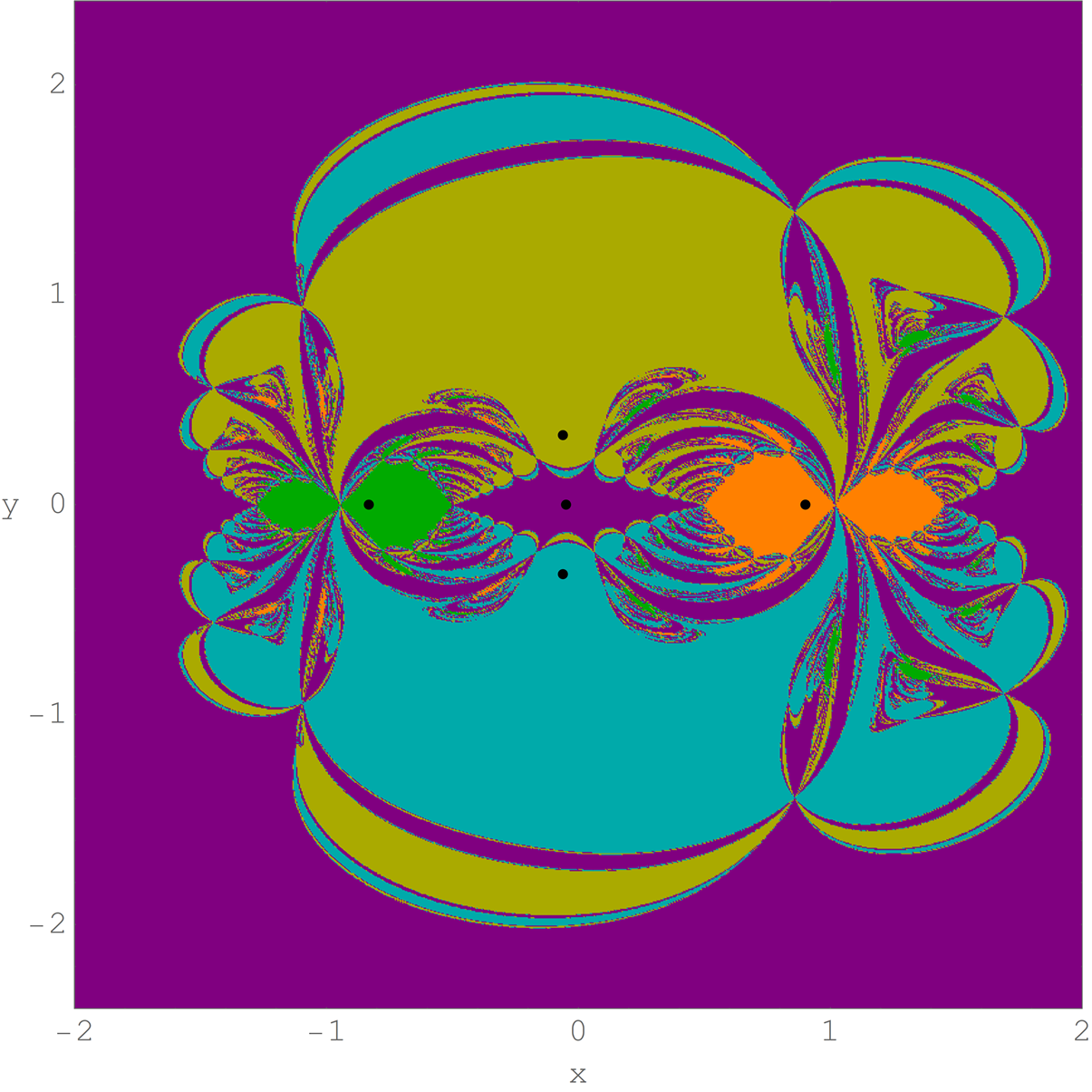}
\includegraphics[scale=0.43]{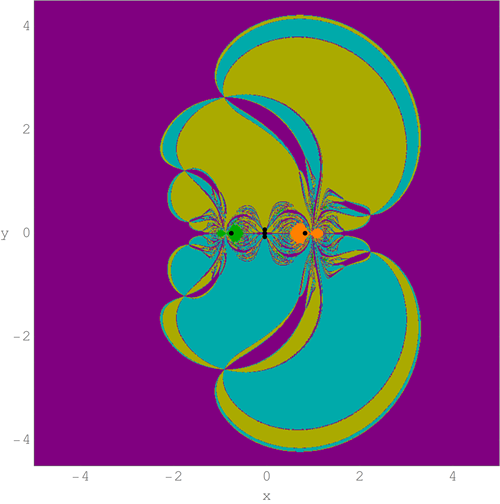}
\caption{The (BoC) linked with the libration points on $(x, y)$-plane  for  $\mu=0.5, q_1=0.15, q_2=0.25$, and then the permissible range is $0.0309757<\omega<1.25144$:   (a)\emph{top left:} for $\omega=0.0309757 + 0.001,$ (b)\emph{top right:} for $\omega=0.1034199519$, (c)\emph{middle left:} for $\omega=0.25$, (d)\emph{middle right:} for $\omega=0.375$,  (e)\emph{bottom left:} for $\omega=0.95$, (f)\emph{bottom right:} for $\omega=1.2332376089$. The dots show the positions of libration points.  (colour figure online).}
\label{Fig:Basin_1}
\end{figure*}
%\subsection{Case:1}

The numerical analysis with the Copenhagen case where the  mass ratio $\mu=0.5$ and for varying values of the angular velocity  are illustrated whereas the value of radiation parameters $q_1=0.15$, $q_2=0.25$.  We start our analysis with Fig. \ref{Fig:Basin_1}a, which is depicted for $\omega=0.0309757 + 0.001$, very close to the critical value. We can observe that the domain of the BoC linked to the libration points is well formed and majority of the area of the finite domains are composed of the mixtures of various types of initial conditions whose final state is unpredictable. Consequently, these areas turn into the chaotic sea. Further, it is noticed that the large number of initial conditions (i.e., $45.25\%$ of the considered initial conditions) converges to the libration point $L_1$, which has infinite extent as well. Whereas $17.12\%$ of considered initial conditions converge to the $L_{4,5}$ and $4.66\%$ of initial conditions converge to the libration point $L_3$. The majority of the area of the finite domain of the BoC is occupied by those initial conditions which either converge to one of the non-collinear libration points.  In Fig. \ref{Fig:Basin_1}b, when the $\omega=0.1034199519$, there exist five equilibrium points and the domain of the BoC linked to the equilibrium points $L_{3}$ and $L_{2}$ resembles to shape of exotic bugs with many legs and antennas which are separated by the chaotic strip composed of various type of initial conditions whose final states are not same. However, the entire $xy$-plane is covered by the well formed BoC. The extent of the domain of BoC linked to the equilibrium point $L_{1}$ is infinite on the other hand for all other equilibrium points these extents are finite. The domain of the  BoC linked to the libration points $L_{4}$ and $L_{5}$ looks like butterfly wings whose wings boundaries are segregated by chaotic mixture of  various types of initial conditions whose final state are different. It is observed that $6.87\%$ of considered initial conditions are converging to the equilibrium point $L_{3}$,  $9.6\%$ of initial conditions are converging to $L_{2}$ whereas $11.74\%$ of initial conditions are converging to each of equilibrium points $L_{4}$ and $L_{5}$ and remaining are converging to $L_{1}$ which has infinite extent.  Further, when the value of $\omega$ increases, the domain of BoC associated to the equilibrium points shrinks significantly except the domain of BoC linked to the libration point $L_{1}$ which consequently increases. Moreover, with the increase in value of angular velocity, the domain of the BoC linked to $L_{4}$ and $L_{5}$  becomes more regular but decreases.   Further,  $3.09\%$ of considered initial conditions converge to the libration point $L_{3}$,  $3.79\%$ of initial conditions converge to libration point $L_{2}$ whereas $11.22\%$ of initial conditions converge to each of libration points $L_{4}$ and $L_{5}$ and $69.1\%$ of initial conditions converge to $L_{1}$ when $\omega= 0.375$.
The finite domain of the BoC continue to decrease  with the increase in value of angular velocity and consequently the infinite domain of BoC increases. It is observed that the most of the finite region of the BoC is covered by the domain of the BoC  associated with the non-collinear in-plane equilibrium points whereas the BoC linked to collinear equilibrium points $L_{2,3}$ look like a very small  bugs without legs and antenna (see  Fig. \ref{Fig:Basin_1}f ) when $\omega= 1.233237608897815$,  only $0.16\%$ of the  initial conditions converge to the libration point $L_{3}$,  $0.22\%$ of initial conditions converge to $L_{2}$ whereas $18.55\%$ of initial conditions converge to each of libration points $L_{4}$ and $L_{5}$ which is slightly higher than the previous cases and remaining are converging to $L_{1}$ with infinite extent. It can be seen that each of the initial conditions converge to one of the attractors sooner or later.
\begin{figure*}
\centering
\includegraphics[scale=0.22]{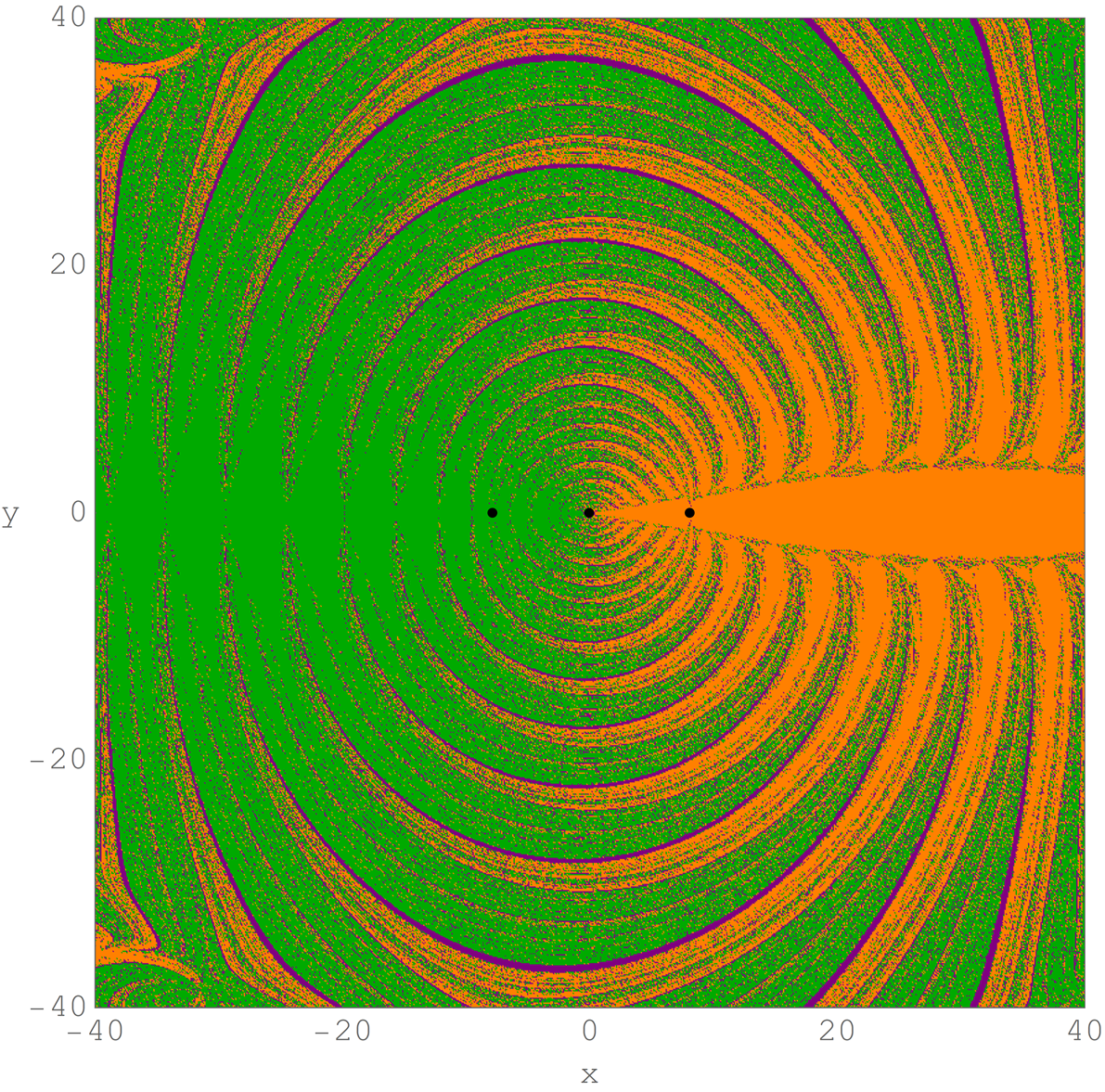}%\omega_0.02
\includegraphics[scale=0.525]{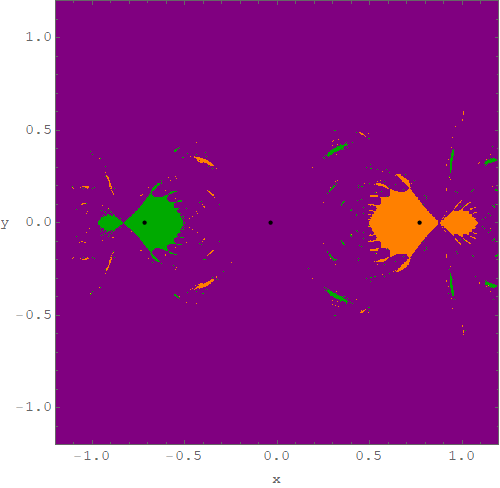}\\%\omega_1.5
\includegraphics[scale=0.525]{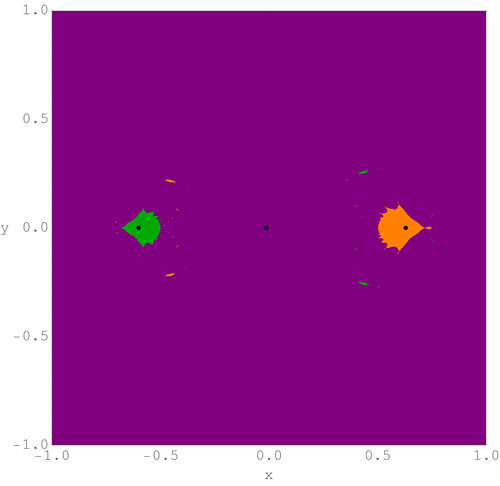}\\%\omega_3.5
\caption{The (BoC) linked with the libration points on $(x, y)$-plane  for  $\mu=0.5, q_1=0.15, q_2=0.25$, and  (a)\emph{top left:} for $\omega=0.02$, (b)\emph{top right:} for $\omega=1.5$, (c)\emph{bottom:} for $\omega=3.5$. The dots show the  positions of libration points.  (colour figure online).}
\label{Fig:Basin_2}
\end{figure*}
%%%%%%%%%%%%%%%%%%%%%%%%%%%%%%%%%%
%%%%%%%%%%%%%%%%%%%%%%%%%%%%%%%%%%

In Fig.\ref{Fig:Basin_2}, the domain of BoC is depicted for those values of $\omega$ for which there exist only collinear libration points, i.e., when $0< \omega <0.0309757$ or $\omega >1.25144$. When $\omega=0.02$ (see Fig.\ref{Fig:Basin_2}a) it is observed that the extent of BoC corresponding to each of libration points looks infinite. We believe that this happens since the value of $\omega$ is very close to zero. It is seen that $53.61\%$ of initial conditions converge to the libration point $L_{3}$ whereas $14.1\%$ and $32.28\%$ of the investigated initial conditions converge to $L_{1}$ and $L_{2}$ respectively. It is noticed that in Fig. \ref{Fig:Basin_2}(b, c) the domain of BoC  linked to libration point $L_{1}$ has infinite extent and for remaining equilibrium points the domain of  BoC are finite.  However, in this case when three libration points exist, it is seen that for $\omega=1.5$ (Fig.\ref{Fig:Basin_2}b) only $1.6\%$ and $2.23\%$  of initial conditions converge to collinear libration points $L_{3}$ and  $L_{2}$ respectively and rest of initial conditions converge to $L_{1}$ which has infinite extent. Further, when $\omega=3.5$ (Fig.\ref{Fig:Basin_2}c) only $0.459\%$ and $0.671\%$  of total considered initial conditions finally enticed by the collinear libration points $L_{3}$ and  $L_{2}$, which unveil the fact that the domain of the BoC linked to these libration points reduces as value of $\omega$ increases.

Indeed, it is very remarkable to compare the Fig.\ref{Fig:Basin_2}a and Fig.\ref{Fig:Basin_1}a, where the number of libration points changes from three to five, respectively. It can be noticed that when value of $\omega$ is just above zero, the domain of the BoC linked to the libration point $L_3$ (see Fig.\ref{Fig:Basin_2}a) looks like antennas of the exotic bugs shaped region, constitutes the domain of the BoCs linked to the libration points $L_{4,5}$ when the value of $\omega$ increases slightly from the critical value (see Fig.\ref{Fig:Basin_2}a). This happens, since the non-collinear libration points just originate in the vicinity of the $L_3$ at the critical value of $\omega$. Further, when we compare Fig.\ref{Fig:Basin_1}f and Fig.\ref{Fig:Basin_2}b, it can be noticed that the domain of the BoC linked to non-collinear libration points $L_{4,5}$ shrinks to the BoC linked to the collinear libration point $L_1$ when the value of $\omega$ crosses the critical value. The main reason for this is the libration points $L_{4,5}$ annihilate in the vicinity of $L_1$ at the critical value of the $\omega$.
%%%%%%%%%%%%%%%%%%%%%%%%%%%%%%%%%%
%%%%%%%%%%%%%%%%%%%%%%%%%%%%%%%%%%
\begin{figure*}
\centering
\includegraphics[scale=0.54]{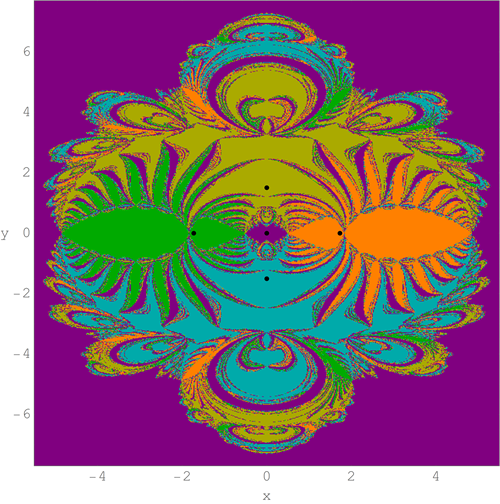}%\omega_1.5
\includegraphics[scale=0.55]{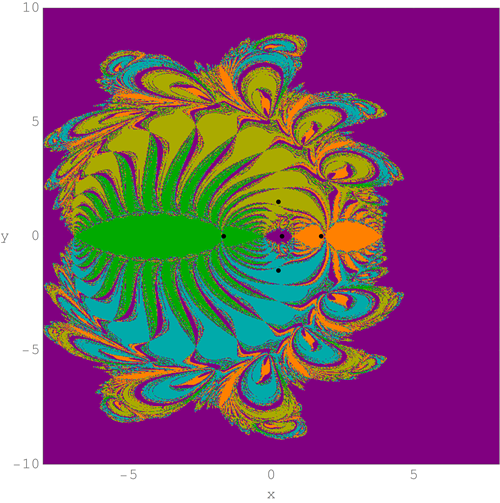}
\includegraphics[scale=0.55]{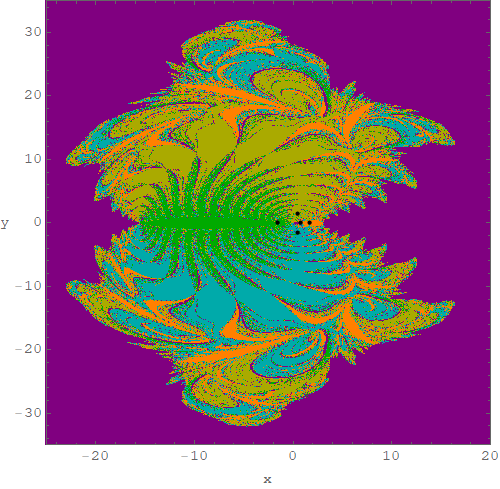}\\
\caption{The (BoC) linked with the libration points on $(x, y)$-plane  for  $\omega=0.5, q_1=q_2=1$, and (a)\emph{top left:} for $\mu=0.5$, (b)\emph{top right:} for $\mu=0.25$, (c)\emph{bottom:} for $\mu=0.05$. The dots show the  positions of libration points.  (colour figure online).}
\label{Fig:Basin_3}
\end{figure*}

In Fig. \ref{Fig:Basin_3}, the BoC are depicted for three different values of the mass parameter $\mu$ in the presence of fixed angular velocity $\omega=0.5$ when the primaries are not radiating. When $\mu=0.5$, the domain of BoC  is symmetrical about both the axes and also the BoC have finite extent  connected to all the equilibrium points except $L_{1}$ which has infinite extent. Further, when the value of the mass parameter decreases, the domain of the BoC connected to those libration points which have finite extent, expands. The domain of the BoC connected to the equilibrium points $L_{2, 3}$ appear as exotic bugs having various legs and antennas whereas the domain of BoC linked to non-collinear libration points appear as multiple butterfly wings when the value of mass parameter decreases these wings become larger. Further, only $9.92\%$ and $6.21\%$ of initial conditions converge to the collinear libration points $L_{3}$ and $L_{2}$ respectively, whereas $11.78\%$ of those initial conditions converge to $L_{4, 5}$ and remaining $60.27\%$ of initial conditions converge to $L_{1}$ when $\mu=0.25$.  In Fig. \ref{Fig:Basin_3}c, when $\mu=0.05$, it can be seen that finite domain of the BoC  expand  and consequently the domain of BoC linked to $L_{1}$ decreases. In addition, the domain of BoC associated to $L_{3}$, which looks like exotic bugs, increases significantly.

If we compare Fig.3f in \cite{Z16} with the Fig. \ref{Fig:Basin_3}a which is illustrated for the same value of $\mu=0.5$ but for $\omega=0.5$, we can observe that the well formed domain of the BoC associated to the libration points having finite extent increases significantly. Moreover, the domain of BoC linked to $L_{4,5}$ which was regular becomes more chaotic when $\omega\neq 1$.
Moreover, when $\omega \neq 1$(in particular $<1$), the fractal structure (in the sense explained in Section \ref{Sec:5}) is higher, as also shown in Fig. \ref{Fig:BE1}b. We rather note the same behavior when $\mu$ decreases from $0.5$ to $0$ in both the cases when $\omega=0.5$ and $\omega=1$. Furthermore, the Fig. \ref{Fig:Basin_3} is done with a more refined initial conditions grid with respect to Fig.3f in \cite{Z16}, and this allows better visualization of the fractal structure.

\begin{figure*}
\centering
\includegraphics[scale=0.45]{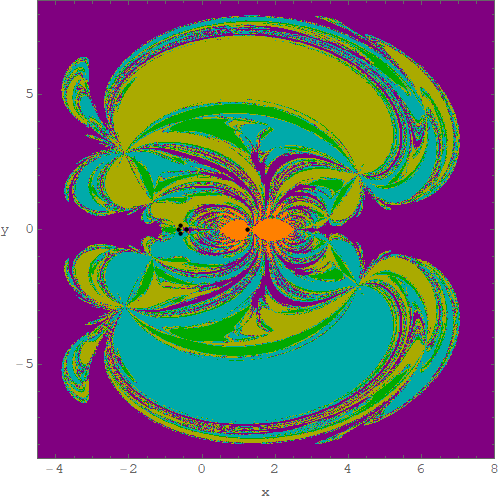}
\includegraphics[scale=0.45]{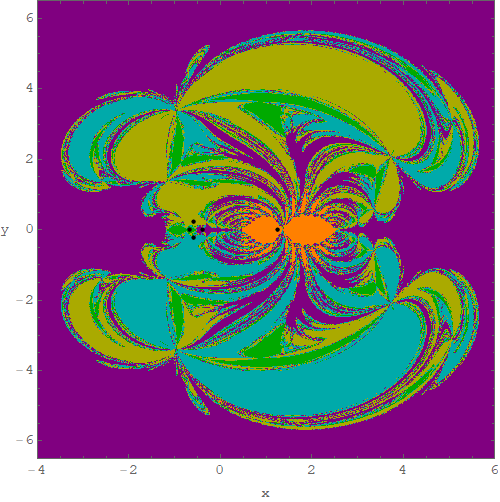}\\
\includegraphics[scale=0.45]{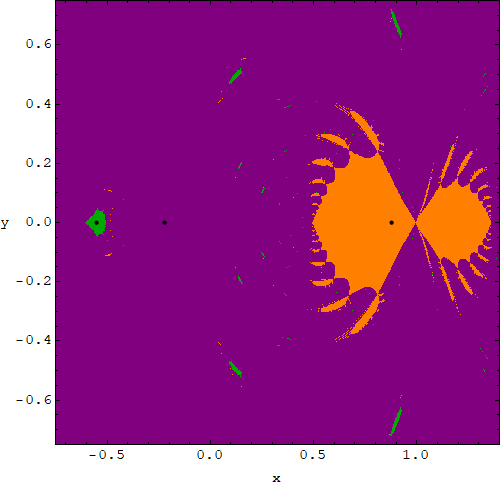}
\includegraphics[scale=0.45]{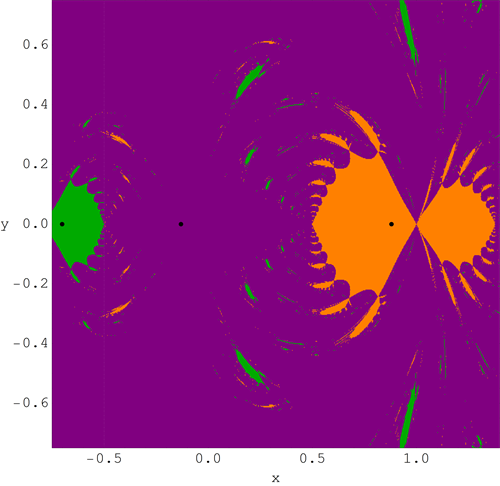}\\
\caption{The BoC linked with the libration points on $(x, y)$-plane.  When  $\omega=0.85, q_2=1, \mu=0.5$, and (a)\emph{top left:} for $q_1=0.004$, (b)\emph{top right:} for $q_1=0.01$. When  $\omega=2, q_2=1$,(c)\emph{bottom left:} for $q_1=0.01$ (d)\emph{bottom right:} for $q_1=0.2$. The dots show the  positions of libration points.(colour figure online).}
\label{Fig:Basin_4N}
\end{figure*}

In Fig. \ref{Fig:Basin_4N}(a,b), the BoC are presented for two different values of radiation parameter $q_1$ and fixed value of $\omega=0.85$. The topology of the domain of BoC associated with the equilibrium points significantly changes with the change in the radiation parameter. In both the cases the domain of BoC linked to libration point $L_1$ has infinite extent. It is noticed that when the value of $q_1$ increases from $0.004$ to $0.01$, the number of initial conditions which converge to $L_1$ has infinite extent, increases from $43.29\%$ to $52.56\%$ and consequently, the area related to finite extent decreases. Moreover, the initial conditions which compose the BoC of the finite extent linked to the non-collinear libration points decrease from $23.85\%$ to $19.83\%$ and the initial conditions which compose the BoC linked to the libration point $L_3$ also decrease whereas for $L_2$, it increase.
In Fig. \ref{Fig:Basin_4N}(c, d), the BoC are presented for two different values of radiation parameter $q_1$ and fixed value of $\omega=2$. We can observe that in this case there exist only three libration points, and the domain of the BoC linked to libration points $L_{2,3}$ increase and consequently, the domain of BoC associated to $L_1$ which has infinite extent decreases with the increase in value of $q_1$.

It is observed that the values of $\omega$ depend on the value of $q_i, i=1,2$, when $q_1=0.004, q_2=1$, the value of $\omega \in (\omega_1=0.771605, \omega_2=1.24732)$ for which the non-collinear libration points exist. Moreover, for Fig. \ref{Fig:Basin_4N}, the value of $\omega=0.85$ is close to the critical value of $\omega$ and for \ref{Fig:Basin_4N}a, where $q_1=0.004$ and for Fig. \ref{Fig:Basin_4N}b, where $q_1=0.01$. Therefore, a comparison with Fig.\ref{Fig:Basin_1}a and Fig. \ref{Fig:Basin_4N}a, which are both illustrated for the very close value of $\omega$ to its critical value, the topology of the domain of BoC in Fig.\ref{Fig:Basin_1}a is very noisy whereas in Fig.\ref{Fig:Basin_4N}a it looks much regular. It is also shown in Fig.\ref{Fig:BE1}b, that as the value of the $\omega$ is small, the value of the basin entropy increases. However, in both the cases the extent of the BoC linked to the central collinear libration point is infinite and for the remaining libration points it is finite.  For Fig. \ref{Fig:Basin_4N}c, where $q_1= 0.01$, we get $\omega\in (\omega_1=0.694922,  \omega_2=1.33999)$ and for \ref{Fig:Basin_4N}d, where $q_1=0.2$, we get $\omega\in (\omega_1=0.267535,  \omega_2=1.99509)$ which shows the interval of $\omega$ for which five libration points exist, consequently in \ref{Fig:Basin_4N}(c,d) the value of $\omega$ is set out of the range so that only three equilibrium points exist. Moreover, \ref{Fig:Basin_4N}d is illustrated for very close value of $\omega$ to its critical value.
We compare Fig. \ref{Fig:Basin_2}(b,c) with \ref{Fig:Basin_4N}(c,d) and observe that the domain of BoC linked to $L_{2,3}$ increases in both cases when $\omega$ increases as well as $q_1$ increases. Further, in all the cases the topology of the BoC are symmetrical about the $x-$axis.

If we compare Fig.9a of Ref.\cite{Z16} (when $\omega=1$) with Fig. \ref{Fig:Basin_4N}b where the value of $\omega=0.85$, we can notice that the basins boundaries are more chaotic in comparison of the previous case moreover, finite regions of the domain of the BoC also increases significantly, however, not much changes are noticed in the topology of the BoC. Also the similar behaviour has been observed for the domain of BoC linked to the libration points $L_1, L_2$, i.e., it increases with the increase in the value of $q_1$. On the contrary, when the value of $\omega\neq1$, the domain of BoC linked to $L_{4,5}$ decreases with the increase in the value of $q_1$.
%%%%%%%%%%%%%%%%%%%%%%%%%%%%%%%%%%
%%%%%%%%%%%%%%%%%%%%%%%%%%%%%%%%%%%%%%%%%%%%%%%%%%%%%%%%%
\section{The Basin Entropy}
\label{Sec:5}
In the analysis of color coded diagrams (CCDs), it is observed that the basin of convergence is highly fractal in the locality of the basins boundaries which unveil the fact that it is quite impossible to judge the final state of the initial conditions  falling inside these fractal regions. The term "fractal"  is simply used in the text to unveil the particular area which shows the fractal-like geometry, without evaluating the fractal dimension (see \cite{AVS01}, \cite{AVS09}). Recently,  a new tool to measure the uncertainty of the basins has been presented in  paper \cite{Daz16},  is named as the “basin entropy” and refers to the geometry of the basins and consequently explore the  concept of unpredictability and fractality in the context of BoC.

The philosophy that works in the background of the method  is to split the phase space into $N$ small cells in which every cell  contains at least one of the total number of final states $N_A$. In addition, the probability to evaluate the state $j$ in the $k-$th cell is denoted by $p_{j,k}$. Using the Gibbs entropy formulae,  the entropy for $j-$th cell is
\begin{equation}\label{Eq:9}
S_j=\sum_{k=1}^{N_A}p_{j,k}\log\Big(\frac{1}{p_{j,k}}\Big).
\end{equation}
The average entropy for the total number of cells $N$ is called as basin entropy, i.e., $S_b$
\begin{equation}\label{Eq:10}
 S_b=\frac{1}{N}\sum_{j=1}^{N}S_j=\frac{1}{N}\sum_{j=1}^{N}\sum_{k=1}^{N_A}p_{j,k}\log\Big(\frac{1}{p_{j,k}}\Big).
\end{equation}
It is necessary to  mention that the result for the basin entropy is highly influenced by the total number of cells $N$, so that a precise value of $S_b$  can be obtained for larger value of $N$. In an attempt to overcome this problem, we use  Monte Carlo procedure to select randomly small cells in the phase space, and we observe that for $N>2\times 10^5$ cells,  the final value of the basin entropy remain unchanged.
%%%%%%%%%%%%%%%%%%%%%%%%%%%%%%%%%%%%%%%%%%%%%
%%%%%%%%%%%%%%%%%%%%%%%%%%%%%%%%%%%%%%%%%%%%%
\begin{figure*}
\centering
\resizebox{\hsize}{!}{\includegraphics{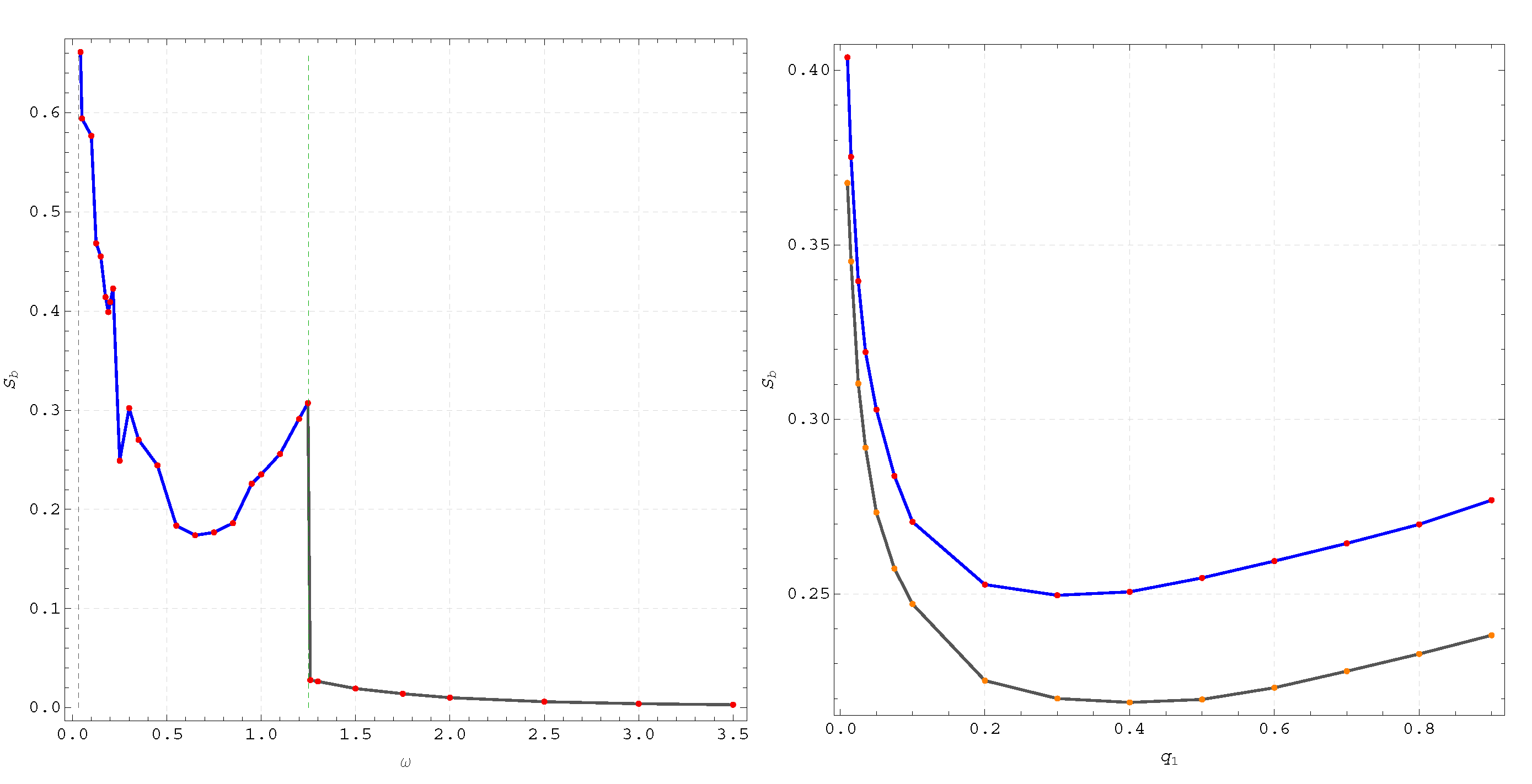}}
\caption{The evolution of the basin entropy $S_b$, of the configuration $(x, y)$ space with $\mu=0.5$: (a)\emph{left:} as a function of the perturbation parameter $\omega$. The vertical, dashed, green lines referred as the value of $\omega$ where the tendency of the parametric evolution of the basin entropy changes as these are the critical value of $\omega$.  (b)\emph{right:} as a function of the perturbation parameter $q_1$ when $q_2=1$. The blue line shows the  basin entropy when the value of $\omega=0.85$ and gray line shows the basins entropy when $\omega=1$.    (colour figure online).}
\label{Fig:BE1}
\end{figure*}
%%%%%%%%%%%%%%%%%%%%%%%%%%%%%%%%%%%%%%%%%%%%%
%%%%%%%%%%%%%%%%%%%%%%%%%%%%%%%%%%%%%%%%%%%%%

In Fig. \ref{Fig:BE1}a, we have illustrated the parametric evolution of the basin entropy for various values of the angular velocity $\omega$, with $\omega\in(0, 3.5)$ when values of radiation parameters are fixed i.e., $q_1=0.15, q_2=0.25$. The gray dashed line shows the value of $\omega\approx0.03097574$, where the value of $S_b$ is maximum. We believe that the value of $S_b$ is maximum as the value of $\omega$ is very close to the critical value. It is further observed that the unpredictability linked to the N-RBoC for the restricted three-body problem is higher when the value of the $\omega\in(0.03097574, 1.25144256)$. However, it started decreasing when $\omega$ increases and at $\omega=0.65$ the value of $S_b=0.1738758105$ is recorded lowest for the all examined value of $\omega$, and again the value $S_b$  increases almost monotonically till $\omega=1.25144256$. The value of  $S_b$ decreases monotonically when $\omega\in(1.25144256, 3.5)$, which refers to the case where only three collinear libration points exist.  Moreover, in the Fig. \ref{Fig:BE1}b, the parametric evolution of the basin entropy is illustrated for the increasing value of the radiation parameter $q_1$ in the both cases i.e., when $\omega=1$ and when $\omega\neq1$.
We observe that the value of the basin entropy remains always higher when the angular velocity $\omega\neq1$. However, the similar tendency in the value of basin entropy has been noticed for the increasing value of the radiation parameter $q_1$ in both the cases.
It is necessary to mention the fact that to illustrate this diagram we have used the numerical results for various additional values of the angular velocity $\omega$ which are not necessarily presented in the Figs. \ref{Fig:Basin_1},\ref{Fig:Basin_2}, \ref{Fig:Basin_3} and \ref{Fig:Basin_4N}.

The main observations can be summarized as follows:
\begin{itemize}
  \item When $\omega\approx1.25144256$ the domain of the BoC and basins boundaries become complicated and consequently the increase in the basin entropy is observed, which is around 0.31. However as the value of the angular velocity increases, the number of the libration points remains three and in this case the value of basin entropy $S_b$ decreases monotonically.
  \item When $\omega= 3.5$ (see Fig. \ref{Fig:Basin_2}c) the value of $S_b\approx0.002981$ which is very close to zero since the topology of the BoC appears very smooth and further increase in $\omega$ shows the same tendency i.e., the smoothness in the basins increases.
  \item When $ 0.0309757<\omega < 1.25144$, there exist five libration points and in this range of $\omega$ it can be observed that the value of the basin entropy changes abruptly. Consequently, for this range of $\omega$ the unpredictability linked to the N-RBoC for the R3BP in the presence of angular velocity $\omega$ is higher.
  \item When the value of radiation parameter $q_1$ increases the value of the basin entropy decreases monotonically when $q_1\in(0,0.4)$ and increases monotonically when $q_1\in(0.4,1)$ and $\omega=1$. Whereas basin entropy decreases monotonically when $q_1\in(0,0.3)$ and increases monotonically when $q_1\in(0.3,1)$ and $\omega=0.85$. It is necessary to note that, although the curves are different, their behaviour are same.
\end{itemize}
%%%%%%%%%%%%%%%%%%%%%%%%%%%%%%%%%%%%%%%%%%%%
%%%%%%%%%%%%%%%%%%%%%%%%%%%%%%%%%%%%%%%%%%%%
\section{Discussion and conclusion}
\label{Sec:6}
In the present paper, we numerically explored the BoC by applying the bivariate version of iterative scheme in the photo-gravitational version of  restricted problem of three bodies when the angular velocity is not equal to unity. The main outcomes of the present study can be summarized as follows:
\begin{itemize}
   \item[*] There exist either five or three libration points for the system.  For fixed values of the $q_i$, $\mu$ and varying values of $\omega$, it can be seen that  the libration point $L_{3}$ moves towards the primary $P_1$ whereas the libration points $L_{1,2}$ move toward the primary $P_2$ as the value of $\omega$ increases. It is observed that the non-collinear libration points originate in the vicinity of the libration point $L_{3}$ at $\omega\approx |q_1^\frac{1}{3}-q_2^\frac{1}{3}|^ \frac{3}{2}$ and these points annihilate in the neighbourhood of the libration point $L_{1}$ when $\omega\approx(q_1^\frac{1}{3}+q_2^\frac{1}{3})^ \frac{3}{2}$.
  \item[*]The attracting domains, linked to the equilibrium point $L_{1}$, extend to infinity, in all studied cases (except for Fig. $\ref{Fig:Basin_2}a$),  while the domain of BoC associated to other libration points are finite. The BoC diagrams, on the configuration $(x, y)$ plane are symmetrical  in all the studied cases, with respect to the horizontal $x$-axis.
  \item[*]The numerical investigations suggest that the multivariate version of Newton-Raphson iterative scheme  converges very fast for those initial conditions which lie in the vicinity of the libration point and converge very slow for those initial conditions which are lying in the vicinity of the basin boundaries. However, all the initial conditions converge to one of the attractors sooner or later.
  \item[*]The numerical investigations unveil that for the interval of $\omega$ where only three libration points exist, the lowest value of $S_b$ is attained near $\omega=3.5$, whereas the highest value of  basin entropy was achieved near $\omega\approx1.25144256$ which is the critical value of $\omega$ where the number of libration points changes. Moreover, for those intervals of $\omega$ in which five libration points exist, the maximum value of the basins entropy $S_b$ is achieved for the value of $\omega\approx0.03097574$, i.e., the starting critical value of $\omega$ when there exit five libration points.   This reveals  the unpredictability, regarding the attracting regions, in the photo-gravitational restricted  three-body problem with angular velocity.
\end{itemize}
In addition, we have used the latest version 12 of Mathematica$^\circledR$ for all the graphical illustrations in this paper. In future, it is worth studying problem by using different  iterative schemes to analyze the similarity as well as difference  on the associated basins of attraction.
%\section*{Acknowledgments}
%\footnotesize
%The authors are thankful to Center for Fundamental Research in Space dynamics and Celestial mechanics (CFRSC), New Delhi, India for providing research facilities.

%The authors would like to express their warmest thanks to the anonymous referee for the careful reading of the manuscript and for all the apt suggestions and comments which allowed us to improve both the quality and the clarity of the paper.

\section*{Compliance with Ethical Standards}
\begin{description}
  \item[-] Funding: The authors state that they have not received any research
grant.
  \item[-] Conflict of interest: The authors declare that they have no conflict of
interest.
\end{description}
\section*{Acknowledgments}
\footnotesize
The authors would like to express their warmest thanks to the anonymous referee for the careful reading of the manuscript and for all the apt suggestions and comments which allowed us to improve both the quality and the clarity of the paper.


\begin{thebibliography}{}
 \bibitem[\protect\citeauthoryear{Aguirre et al.}{2001}]{AVS01} Aguirre, J., Vallejo, J.C., Sanju´an, M.A.F.: Wada basins and chaotic invariant sets in the \'{H}enon-Heiles system,\emph{ Phys. Rev. E} 64 (2001) 066208. \url{https://doi.org/10.1103/PhysRevE.64.066208}
\bibitem[\protect\citeauthoryear{Aguirre et al.}{2009}]{AVS09}Aguirre, J., Viana, R.L., Sanju´an, M.A.F.: Fractal Structures in nonlinear dynamics, \emph{Rev. Mod. Phys. }81 (2009) 333-386. \url{https://doi.org/10.1103/RevModPhys.81.333}

\bibitem[\protect\citeauthoryear{Abouelmagd and Abdullah }{2019a}]{AA19a}Abouelmagd, E.I., Abdullah A. A.: The motion properties of the infinitesimal body in the framework of bicircular Sun perturbed Earth-Moon system. \emph{New Astronomy},
Volume 73, November 2019a, 101282. \url{https://doi.org/10.1016/j.newast.2019.101282}
	
\bibitem[\protect\citeauthoryear{Abouelmagd et al.}{2019b}]{AGL19b}Abouelmagd, E.I., Garcia Guirao, J.L., Llibre, J.: Periodic orbits for the perturbed planar circular restricted 3-body problem.  \emph{Discrete and Continuous Dynamical Systems - Series B}, 24 (3), pp. 1007-1020, (2019b). \url{http://aimsciences.org//article/doi/10.3934/dcdsb.2019003}
\bibitem[\protect\citeauthoryear{Abouelmagd}{2012}]{A12}Abouelmagd, E.I.: Existence and stability of triangular points in the restricted three-body problem with numerical applications. \emph{Astrophysics and Space Science}, 342 (1), pp. 45-53, (2012). \url{https://link.springer.com/article/10.1007\%2Fs10509-012-1162-y}

\bibitem[\protect\citeauthoryear{Abouelmagd et al.}{2016}]{AAG09}Abouelmagd E.I., Alzahrani F., Guiro J.L.G., Hobiny A.: Periodic orbits around the collinear libration points. J. Nonlinear Sci. Appl. (JNSA). (2016) 9 (4): 1716 -1727. \url{http://www.emis.de/journals/TJNSA/includes/files/articles/Vol9_Iss4_1716--1727_Periodic_orbits_around_the_collinea.pdf}

\bibitem[\protect\citeauthoryear{Alzahrani et al.}{2017}]{AAG17}Alzahrani F., Abouelmagd E. I., Guirao J. L.G., Hobiny A.:  On the libration collinear points in the restricted three--body problem. Open Physics (2017) 15 (1): 58 – 67. \url{https://doi.org/10.1515/phys-2017-0007}
\bibitem[\protect\citeauthoryear{Baltagiannis and Papadakis}{2011}]{BP11}Baltagiannis, A.N., Papadakis, K.E.: Equilibrium points and
their stability in the restricted four-body problem,\emph{ Int. J. Bifurc. Chaos }21 (2011) 2179-2193. \url{https://www.worldscientific.com/doi/abs/10.1142/S0218127411029707}

\bibitem[\protect\citeauthoryear{Chernikov}{1970}]{C70} Chernikov, Y. A.: The Photogravitational Restricted Three-Body Problem,  \emph{Soviet Astronomy}, Vol. 14, p. 176, 1970.\url{http://adsabs.harvard.edu/full/1970SvA....14..176C}

\bibitem[\protect\citeauthoryear{Chermnykh}{1987}]{C87} Chermnykh, S. V.: Stability of libration points in a gravitational
field,  \emph{Leningradskii Universitet Vestnik Matematika Mekhanika Astronomiia}, vol. 2, pp. 73--77, 1987.
\bibitem[\protect\citeauthoryear{Daza et al.}{2016}]{Daz16}Daza, A., Wagemakers, A., Georgeot, B., G\'{a}ery-Odelin,
D.,  Sanj\'{a}an, M.A.F.: “Basin entropy: A new
tool to analyze uncertainty in dynamical systems,”
Scient. Rep. 6, 31416(2016). \url{https://www.nature.com/articles/srep31416}
\bibitem[\protect\citeauthoryear{Douskos}{2010}]{D10}Douskos, C.N.: Collinear equilibrium points of Hill's problem with radiation and oblateness and their fractal basins of attraction, \emph{Astrophys. Space Sci.} 326 (2010) 263--271. \url{https://link.springer.com/article/10.1007/s10509-009-0213-5}
\bibitem[\protect\citeauthoryear{ Go\'{z}dziewski and Maciejewski}{1987}]{CM87} Go\'{z}dziewski, K.,  Maciejewski, A. J.: Nonlinear stability of the Lagrangian libration points in the Chermnykh problem, \emph{Celestial Mechanics and Dynamical Astronomy}, vol. 70, no. 1, pp.41--58, 1998. \url{https://link.springer.com/article/10.1023/A:1008250207046}
\bibitem[\protect\citeauthoryear{Pathak et al.}{2015}]{PAT19} Pathak N., Abouelmagd E.I., Thomas V.O.:  On higher order of resonant periodic orbits in the photogravitational restricted three body problem. \emph{The Journal of the Astronautical Sciences}, 66(4), 475-505(2019). \url{https://link.springer.com/article/10.1007/s40295-019-00178-z}
\bibitem[\protect\citeauthoryear{Perdios et al.}{2015}]{PK15} Perdios, E.A., Kalantonis, V.S., Perdiou, A.E.,  and Nikaki, A. A.: Equilibrium points and related periodic motions in the restricted three-body problem with angular velocity and radiation effects. \emph{ Advances in Astronomy}
Volume 2015, Article ID 473483(2015), \url{http://dx.doi.org/10.1155/2015/473483}
\bibitem[\protect\citeauthoryear{Perdios and Ragos}{2004}]{PR04}Perdios, E.A., Ragos, O.: Asymptotic and periodic motion
around collinear equilibria in Chermnykh's problem, \emph{Astronomy and Astrophysics}, vol. 414, no. 1, pp. 361--371, 2004. \url{DOI: 10.1051/0004-6361:20041216}
\bibitem[\protect\citeauthoryear{Prosmiti et al}{1996}]{PFG96}Prosmiti,  R., Farantos, S.C.,  Guantes, R.,  Borondo, F., Benito, R.M.: A periodic orbit analysis of the vibrationally highly excited LiNC/LiCN: a comparison with quantum mechanics,” \emph{The Journal of Chemical Physics}, vol. 104, no. 8, pp. 2921--2931,
1996. \url{https://doi.org/10.1063/1.471113}
\bibitem[\protect\citeauthoryear{Sano}{2007}]{S07} Sano, M. M.: Dynamics starting from zero velocities in the classical Coulomb three-body problem,\emph{ Physical Review E}, vol. 75, no. 2, Article ID 026203, 2007. \url{https://journals.aps.org/pre/abstract/10.1103/PhysRevE.75.026203}
\bibitem[\protect\citeauthoryear{Selim et al.}{2019}]{SGA19}Selim H. H., Guirao J. L.G., Abouelmagd E I.:  Libration points in the restricted three-body problem: Euler angles, existence and stability. \emph{Discrete and Continuous Dynamical Systems}--Series S (DCDS-S) (2019) 12 (4\&5): 703--710. \url{https://www.aimsciences.org/article/doi/10.3934/dcdss.2019044}
\bibitem[\protect\citeauthoryear{ Suraj et al.}{2017a}]{SAA17}Suraj, M.S., Aggarwal, R., Arora, M.: On the restricted four-body
problem with the effect of small perturbations in the Coriolis and centrifugal forces, \emph{Astrophys. Space Sci}. 362 (2017) 159. \url{https://link.springer.com/article/10.1007/s10509-017-3123-y}
\bibitem[\protect\citeauthoryear{ Suraj et al.}{2017b}]{SAP17}Suraj, M.S., Asique, M.C., Prasad, U., Hassan, M.R., Shalini, K.: Fractal basins of attraction in the restricted four-body problem when the primaries are triaxial rigid bodies, \emph{Astrophys. Space Sci.}
362 (2017) 211. \url{https://link.springer.com/article/10.1007/s10509-017-3188-7}
\bibitem[\protect\citeauthoryear{Suraj et al.}{2019}]{sur19} Suraj, M.S., Sachan, P., Zotos, E.E., Mittal. A., Aggarwal, R.: On the fractal basins of convergence of the libration points in the axisymmetric five-body problem: The convex configuration, \emph{Int. J.  of Non-Linear Mechanics,} \textbf{109} (2019) 80-106. \url{https://www.sciencedirect.com/science/article/pii/S0020746218303111}
\bibitem[\protect\citeauthoryear{Suraj et al.}{2019}]{sur19b} Suraj, M.S., Sachan, P., Zotos, E.E., Mittal. A., Aggarwal, R.: On the Newton-Raphson basins of convergence associated with the libration points in the axisymmetric restricted five-body problem: The concave configuration, \emph{Int. J.  of Non-Linear Mechanics,} \textbf{112} (2019) 25-47. \url{https://www.sciencedirect.com/science/article/pii/S0020746218306942}
\bibitem[\protect\citeauthoryear{ Suraj et al.}{2019}]{SAR19}Suraj, M.S., Abouelmagd, E.I., Aggarwal, R.,  Mittal, A.: The analysis of restricted five-body problem within frame of variable mass. \emph{New Astronomy}
Volume 70, July 2019, Pages 12-21. \url{https://www.sciencedirect.com/science/article/pii/S1384107618303245}
\bibitem[\protect\citeauthoryear{Suraj et al.}{2019}]{Sur19d} Suraj, M.S.,    Sachan, P.,   Mittal, A., Aggarwal, R., The effect of small perturbations in the Coriolis and centrifugal forces in the axisymmetric restricted five-body problem, \emph{Astrophys Space Sci} (2019) 364: 44. \url{https://doi.org/10.1007/s10509-019-3528-x}
\bibitem[\protect\citeauthoryear{Suraj et al.}{2019d}]{SUR20}Suraj,M.S.,  Aggarwal, R.,  Mittal,A.,  Meena,O.P., Asique, M.C.:
On the spatial collinear restricted four-body problem with non-spherical primaries. \emph{Chaos, Solitons \& Fractals}, 133(2020)109609. \url{https://doi.org/10.1016/j.chaos.2020.109609}

\bibitem[\protect\citeauthoryear{ Zotos}{2016}]{Z16}Zotos, E.E.: Fractal basins of attraction in the planar circular restricted three-body problem with oblateness and radiation pressure, \emph{Astrophys. Space Sci.} 361 (2016) 181. \url{https://link.springer.com/article/10.1007/s10509-016-2769-1}
\bibitem[\protect\citeauthoryear{ Zotos}{2017}]{Z17} Zotos, E.E.: Comparing the fractal basins of attraction in the Hill problem with oblateness and radiation, \emph{Astrophys. Space Sci.} 362 (2017) 190. \url{https://link.springer.com/article/10.1007/s10509-017-3169-x}
\bibitem[\protect\citeauthoryear{ Zotos and Suraj}{2018}]{ZS18} Zotos, E.E., Suraj, M.S.: Basins of attraction of equilibrium points in the planar circular restricted five-body problem, \emph{Astrophys. Space Sci.}363 (2018) 20. \url{https://link.springer.com/article/10.1007/s10509-017-3240-7}
\end{thebibliography}
\end{document}